\documentclass[apj]{emulateapj}
\usepackage{apjfonts}
\usepackage{amsbsy}
\providecommand{\kms}{\,\ensuremath{\rm{km\,s}^{-1}}}
\newcommand{\apll}{\lesssim}

\slugcomment{Accepted for Publication in ApJ Letters}

\shorttitle{Galactic Environment of NeVIII Absorber}
\shortauthors{Mulchaey and Chen}

\begin{document}


\title{The Galactic Environment of
the NeVIII Absorber toward HE0226$-$4110}

\author{John S. Mulchaey} 
\affil{The Observatories of the Carnegie Institution of Washington, 813 Santa Barbara St.,
    Pasadena, CA 91101, U.S.A.}
\email{mulchaey@ociw.edu}

\and

\author{Hsiao-Wen Chen}
\affil{Department of Astronomy and Astrophysics and Kavli Institute for Cosmological Physics, University of Chicago, Chicago, IL, 60637,U.S.A.}
\email{hchen@oddjob.uchicago.edu}

\begin{abstract}
We report the discovery of a small galaxy system in the vicinity of
the NeVIII absorber at $z=0.20701$ toward HE0226$-$4110. The galaxy
system consists of two 0.25 L$_{*}$ disk galaxies and a 0.05 L$_{*}$
galaxy all within $\Delta\,v < 300$ \kms\ and $\rho\le 200\
h^{-1}$ physical kpc of the absorber.  We consider various scenarios
for the origin of the NeVIII absorption, including photo-ionized gas
from a AGN, a starburst driven wind, a hot intragroup medium, hot gas
in a galaxy halo, and a conductive front produced by cool clouds
moving at high speed through a hot medium.  
We argue that the
conductive front scenario is most likely responsible for producing the
NeVIII feature, because it is consistent with the observed galactic
environment around the absorber and because it naturally explains the
multi-phase nature of the gas and the kinematic signatures of the
absorption profiles.
Although our preferred scenario suggests that NeVIII may not be
directly probing the warm-hot intergalactic medium, it does imply the existence
of an extended hot confining medium around a disk galaxy 
that may contain a significant reservoir of baryons in the form of hot gas.
\end{abstract}


\section{Introduction}

A census of the observed baryons in the local universe falls far short
of the amount required by Big Bang Nucleosynthesis and the cosmic
microwave background \citep{fukugita98}. Cosmological simulations
suggest most of the missing baryons are in the form of 10$^5$--10$^7$K
gas that resides in low-density regions such as groups and
filaments \citep{co99,dave01}.  Considerable effort has been made in
recent years to locate this \lq\lq Warm-Hot Intergalactic Medium\rq\rq
\ (WHIM) \citep{bregman07}, but such gas has largely eluded
detection.  Ultraviolet absorption lines in the spectra of background
quasars are considered a sensitive probe of the WHIM
\citep{verner94,m96}.  The most commonly used feature is the OVI
$\lambda$$\lambda$1031.926,1037.617 \AA\ absorption doublet.  While
many OVI systems have now been identified
\citep{tripp00,tripp08,savage02,ds05,tc08a,tc08b}, recent observations
\citep{tripp08,tc08b} suggest that 
$\sim$30--50\% of the observed OVI absorbers may result from
photo-ionized gas and therefore may not be probing the WHIM.

The presence of other high ionization lines may help distinguish
between photo-ionized and collisionally-ionized systems.  In
particular, the ultraviolet lines of NeVIII
$\lambda$$\lambda$770.5,780.3 are potentially powerful tracers of hot
collisionally ionized gas because of the high ionization potential
required to produce the ions (207.28 eV). 
However, only one intergalactic NeVIII absorber has been reported to
date \citep{savage05,l06}. The observed NeVIII absorption is 
accompanied by lower ionization features.
A detailed ionization
analysis shows that the 
NeVIII is best explained by
collisionally ionized warm gas, while the low ionization features most likely 
result from photo-ionization
\citep{savage05}. The relationship between the collisionally-ionized and photo-ionized 
components is unclear. The galactic environment of the NeVIII absorber is also unknown. 
Here, we report the results of a galaxy
survey in the vicinity of the NeVIII absorber that 
allows us to better understand the nature of this system and to examine
the effectiveness of NeVIII as a probe of the WHIM.  We adopt a
$\Lambda$CDM cosmology, $\Omega$$_{\rm M}$ = 0.3 and
$\Omega$$_{\Lambda}$= 0.7, with a dimensionless Hubble constant h =
H$_{0}$/(100 Mpc s$^{-1}$ km$^{-1}$) throughout this {\it Letter}.

\section{The NeVIII Absorber at $z=0.20701$ Toward HE0226$-$4110}


Using high-resolution FUSE and STIS observations with respective
spectral resolutions of FWHM$\approx 25$ \kms\ and 7 \kms,
\citet{savage05} reported the first detection of intergalactic
NeVIII $\lambda\lambda770,780$ absorption at
$z=0.20701$ toward HE0226$-$4110, together with detections of
absorption transitions due to HI, CIII, OIII, OIV, OVI,
NIII, SiIII, SVI and possibly SV.  Different ionization
species exhibit different kinematic signatures in the absorption-line
profiles, suggesting a multi-phase medium nature of the underlying
absorbing gas.  Specifically, the absorption transitions due to
low-ionization species such as HI, CIII, and OIII are resolved
into two relatively narrow components (Dopper parameter $b\approx 15$
\kms) separated by $\Delta\,v\approx 30$ \kms\ (see also Thom \& Chen
2008b).  The observed HI column densities of the absorbing gas are
$\log\,N({\rm HI})=15.06\pm 0.04$ for the blueshifted component and
$\log\,N({\rm HI})=14.89\pm 0.05$ for the redshifted component.  The
OVI absorption doublet shows a broad component of $b_{\rm
  OVI}\approx 31\pm 2$ \kms\ that centers roughly between the two
components found in the low-ionization lines with $\log\,N({\rm
  OVI})=14.37\pm 0.03$, while the NeVIII (and SVI) absorption
lines appear to be blueshifted by $\approx 7-18$ \kms\ with respect to
the OVI absorption doublet with $\log\,N({\rm NeVIII})=13.9\pm
0.1$ ($\log\,N({\rm SVI})=12.8\pm 0.1$) and $b_{\rm NeVIII}=23\pm
15$ \kms\ ($b_{\rm SVI}\approx 39\pm 20$ \kms).

\citet{savage05}
conducted an extensive ionization analysis of the absorbing gas.
Using the observed absorption strengths of CIII, OIII, NIII, and
SiIII and assuming a mean background radiation field of
$J_{912}=1.9\times 10^{-23}$ ergs cm$^{-2}$ s$^{-1}$ Hz$^{-1}$
sr$^{-1}$ at $z=0.2$ from \citet{hm96}, the authors found
that the relative abundances of these low-ionization species are 
characterized by a photo-ionized cloud of metallicity $\approx 1/3$
solar, temperature $T\sim 2\times 10^4$ K, density $n_{\rm
  H}=2.6\times 10^{-5}$ cm$^{-3}$, neutral fraction $f_{\rm H^0} =
2.5\times 10^{-4}$, and size $\approx 57$ kpc.  But applying a
photo-ionization model to explain the observed large column densities
of OVI and NeVIII with the same assumed radiation field 
would require the gas density to be still
lower, $n_{\rm H}=4.5\times 10^{-7}$ cm$^{-3}$ and the inferred total
pathlength near $\sim\,11$ Mpc.  Such a scenario was ruled out by
\citet{savage05}, because the Hubble flow broadening of the absorption
over such a large path length would be much larger than that found in
the observed line widths.

\citet{savage05} further considered a collisional ionization scenario,
adopting the $1/3$ solar metallicity inferred for the low-ionization
species under the photo-ionization model.  The authors derived a gas
temperature of $T = 5.4 \times 10^{5}$ K, assuming the gas is under
collisional ionization equilibrium.  At this temperature, the observed
line widths of NeVIII and OVI would imply a turbulent velocity
field of $b_{\rm turbulent}\sim 20$ \kms.  At $T\apll 6\times 10^5$ K,
however, cooling efficiency approaches a maximum.  The gas is
therefore expected to be in non-equilibrium conditions, undergoing
rapid cooling at somewhat lower temperature (e.g.\ Gnat \&
Sternberg 2007).  

In summary, \citet{savage05} concluded that the $z=0.20701$ absorber
toward HE0226$-$4110 is part of a multi-phase system.  In their
preferred model, the low ionization features originate in a
photo-ionized component of $T\sim 2\times 10^4$ K, while the NeVIII
and most of the OVI absorption originate in collisionally ionized
gas of $T\sim 5\times 10^5$ K.  
However, the connection between the
photo-ionized cool component and the collisionally ionized hot gas is
unclear.

\section{The Galaxy System at $z=0.207$}

\begin{deluxetable*}{crrrrcccccc}
\tabletypesize{\tiny}
\tablecaption{Summary of Galaxies Associated with the NeVIII Absorber at $z=0.20701$ toward HE0226$-$4110}
\tablewidth{0pt}
\tablehead{\colhead{Galaxy} & \colhead{$\Delta\alpha$} & \colhead{$\Delta\,\delta$} & \colhead{$\Delta\,\theta$} & \colhead{$\rho$} & \colhead{} & \colhead{} & \colhead{} & \colhead{} & \colhead{Spectral} & \colhead{$M_R$} \\
\colhead{ID} & \colhead{(arcsec)} & \colhead{(arcsec)} & \colhead{(arcsec)} & \colhead{($h^{-1}$ kpc)} & \colhead{$B$} & \colhead{$R$} & \colhead{$I$} & \colhead{$z_{\rm spec}$} & \colhead{Type$^a$} & \colhead{$-5\,\log\,h$} }
\startdata
$A$  &  $-7.9$ & $-7.9$  &  11.2  & 26.5  & $22.85\pm 0.06$ &  $21.94\pm 0.04$ & $21.20\pm 0.03$ & 0.2065 &  2 & $-17.2$ \nl
$B$  & $-19.8$ & $-25.3$ &  32.1  & 76.5  & $22.05\pm 0.03$ &  $20.29\pm 0.01$ & $19.38\pm 0.01$ & 0.2078 &  2 & $-18.9$ \nl
$C$  & $-53.4$ & $63.5$  &  82.9  & 197.3 & $22.71\pm 0.05$ &  $20.53\pm 0.01$ & $19.65\pm 0.01$ & 0.2077 &  1 & $-18.9$ \nl
\enddata
\tablenotetext{a}{Spectral type of the galaxies: ``1'' indicates
absorption-line dominated galaxies, and ``2'' indicates emission-line
dominated galaxies.  See also Figure 1.}
\end{deluxetable*}

To help determine the nature of the $z=0.20701$ absorption system, we
have performed a spectroscopic survey of the galaxies in the field
around HE0226$-$4110.  Complete details of our observing program can
be found in \citet{cm09}.  Here we provide a brief summary of the
galaxy survey.  Spectroscopic candidates were selected from optical
images covering a $\sim$ 28 $\times$ 28 arcmin$^2$ region centered on
the QSO.  Candidate galaxies were observed at the Magellan telescopes
using the IMACS and LDSS-3 spectrographs. The spectra were reduced
using the COSMOS software package and redshifts measured by
cross-correlating the extracted spectra and galaxy templates.
The mean redshift measurement uncertainty is $\Delta z$ $\approx$  $0.0003$.
Our survey is nearly 100\% complete for galaxies brighter than $R=23$
at angular distances $\Delta\,\theta\le 2'$ (corresponding to
projected distances of $\rho\le 285\ h^{-1}$ physical kpc at
$z=0.20701$) and approximately 50\% complete at $R\le 22$ out to
$\Delta\,\theta\le 10'$ ($\rho\le 1.4\ h^{-1}$ physical Mpc at
$z=0.20701$).

Our spectroscopic survey revealed a small association of three
galaxies in the vicinity of the NeVIII absorber. These galaxies are
all within $|\Delta\,v| < 300$ \kms\ and $\rho\le 200 h^{-1}$ physical
kpc of the absorber.  A summary of their properties is presented in
Table 1.  The optical spectra and an $R$-band image are presented in
Figure 1.

Table 1 shows that the galaxy with the smallest projected
distance from the QSO ($A$) is an underluminous disk galaxy of
luminosity $\sim 0.05\,L_{*}$ (adopting $M_{R_*}-5\,\log\,h=-20.4$
from Blanton et al.\ 2003).  The projected distance between $A$
and HE0226$-$4110 is $\rho=26.5\,h^{-1}$ physical kpc.  The two
other objects are more luminous disk galaxies of $\sim\,
0.25\,L_{*}$.  These galaxies have impact parameters of $\rho=76.5\
h^{-1}$ physical kpc ($B$) and $\rho=197.3\ h^{-1}$ physical
kpc ($C$), respectively. Galaxies $A$ and $B$ have
emission-line spectra consistent with normal star-forming galaxies,
while $C$ exhibits absorption dominated spectral features that
indicates a more evolved underlying stellar population. 

We note that based on the completeness of our survey, we can rule out
the presence of additional galaxies near the redshift of the absorber
down to $M_R - 5\log\,h = -16.1$ (roughly $0.02\,L_{*}$) within $\rho
= 285\ h^{-1}$ kpc of the QSO.  Four additional galaxies of $M_R -
5\log\,h = -17.2$ to $M_R - 5\log\,h = -20.45$ are found within
$|\Delta\,v| < 300$ \kms\ of the absorber but these galaxies are
located at $\rho=802-1500\ h^{-1}$ physical kpc from the QSO line of
sight (Chen \& Mulchaey 2009).  They are therefore unlikely to host
the absorbing gas found in the QSO absorption-line data.

\section{The Origin of the NeVIII Absorber}

The combination of the results of our galaxy survey in the field
around HE0226$-$4110 and the known properties of the absorbing gas
from Savage et al.\ (2005) provides a unique opportunity to examine
the relationship between the highly ionized warm gas and the galaxy
environment.  Below we consider whether the absorber is
produced in individual galactic halos, intragroup medium in a galaxy
group, or outflows from either a starburst galaxy or AGN.  
The close proximity of the NeVIII absorber to multiple galaxies is
similar to what is found for some OVI absorbers (e.g.\ Stocke et al.\
2006; Wakker \& Savage 2009).  It suggests that the absorbing gas is
unlikely to originate in underdense filamentary regions, but is likely
to be gravitationally bound to galaxies.

\subsection{Intragroup Medium}

To assess whether the absorber is likely to be associated
with an intragroup medium or a halo around a galaxy, we first estimate
individual halo sizes of the galaxies based on their intrinsic
luminosities.  Previous halo occupation studies of field galaxies have
shown a monotonic relationship between halo mass $M_h$ and galaxy
luminosity for galaxies fainter than $L_*$ and located at the centers
of their dark matter halos. Adopting the
$M_h$ vs.\ $M_R$ relation of Tinker \& Conroy (2009), we estimate that
galaxies $A$, $B$, and $C$ reside in halos of $M_h=10^{11}$,
$10^{11.5}$, and $10^{11.5}\ h^{-1}\,M_{\odot}$, respectively.  The
corresponding halo radii $R_h\equiv R_{200}$\footnote{We adopt
$R_{200}$ to represent the size of a dark matter halo, which
corresponds to the radius within which the enclosed mean density is
200 times the background matter density.}  
are $R_h=93$, 137, and $137\ h^{-1}$ kpc for $A$, $B$, $C$,
respectively.  The expected halo radii of these galaxies together with
the projected separations between one another indicate that galaxies
$B$ and $C$ do not share a common halo, whereas galaxies $A$ and $B$
(separated by $\approx 50\ h^{-1}$ projected physical kpc and $\approx
300$ \kms) are likely to share a common halo (based on the two-point 
correlation function,
the probability of having two
unrelated galaxies at this separation is $\sim$ 0.5\%; \citet{z05}).  Analogous
to the Local Group, it is probably more appropriate to consider $B$
and $C$ as two separate halos, instead of a single virialed \lq\lq
group\rq\rq \ . With this interpretation of the mass distribution, it
seems unlikely that the NeVIII absorber would be associated with a
diffuse intragroup medium.

\subsection{Extended Hot Gaseous Halo}

\begin{figure*}
\begin{center}
\includegraphics[scale=0.5]{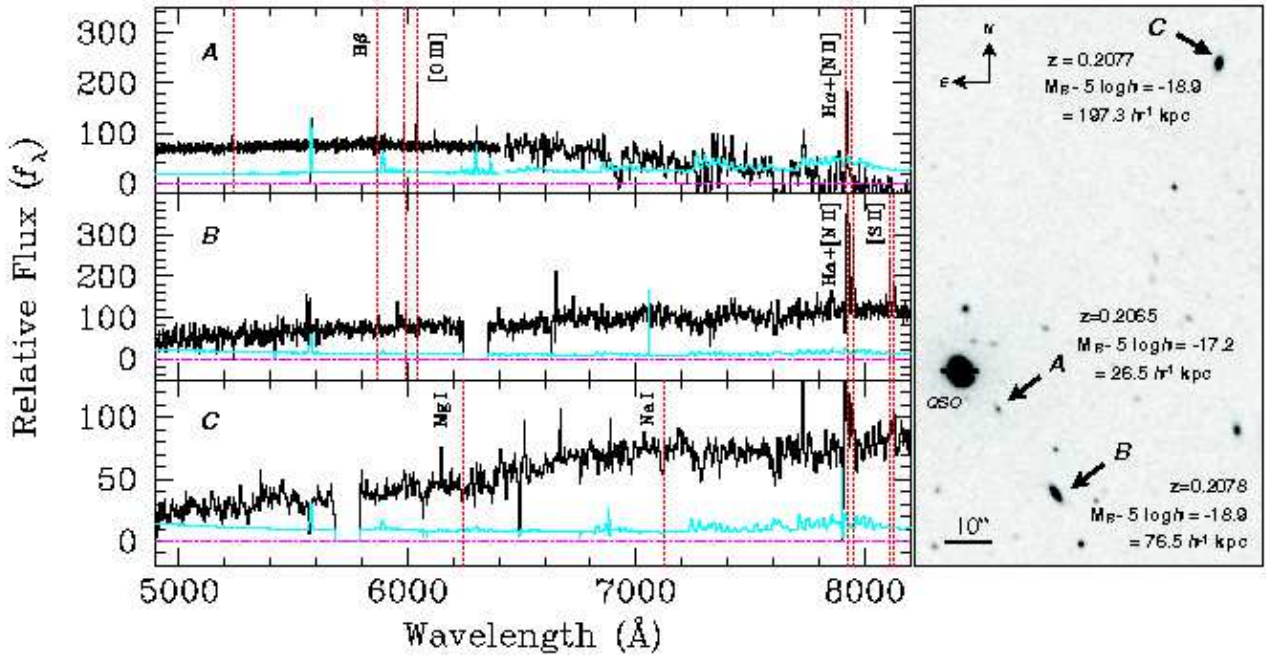}
\caption{{\it Left}: Optical spectra of the three galaxies found
  within projected distances $\rho<200\ h^{-1}$ kpc of the NeVIII
  absorber at $z=0.20701$ toward HE0226$-$4110.  The
  spectra of galaxy $A$ were obtained with LDSS3, while the 
  spectra of galaxies $B$ and $C$ were
  obtained with IMACS (see Chen \& Mulchaey 2009). The system
  response function of each slit setup has been removed based on a
  crude flux calibration using the spectra of alignment stars.
{\it Right}:
  $R$-band image of the field around HE0226$-$4110. 
The seeing was $\sim$ 0.65$''$. 
}
\end{center}
\end{figure*}

Given the projected distances from the QSO and the $R_{200}$ estimates above, 
only $A$ and $B$ are viable hosts for the NeVIII absorber.
Here, we consider the possibility that the absorber originates
in an extended hot halo around one of these galaxies. 
Hot halos around individual galaxies are expected in the hierarchical
galaxy formation paradigm.  
While such halos are
commonly found around early-type galaxies (Fabbiano et al.
1992), the presence of extended hot halos around disk galaxies has not
been confirmed (Rasmussen et al. 2009) with the exception of
the Milky Way for which an extended hot halo is implied from
several observations \citep{
sembach03,fang07,bl07}.

We first consider the more massive galaxy $B$.
The estimated halo mass of galaxy $B$ 
would imply a virial temperature of $T_{\rm
vir}\sim 4\times 10^5$ K, similar to the gas temperature ($T=
5.4\times 10^5$ K) derived for the NeVIII absorber under the
assumption of a collisional ionization equilibrium (Savage et al.\
2005).  It is therefore possible that the absorber originates in an
extended hot halo around galaxy $B$.

If the NeVIII absorption is produced by an extended hot halo, we can
use the observed properties of the absorption to estimate the mass of
the hot gas.  \citet{savage05} derive a total hydrogen column density
of $\log N({\rm H}) = 19.9$ for the absorber under the assumptions of
collisional ionization equilibrium and metallicity [Z/H]$=-0.5$.
Assuming a gaseous radius equal to $R_{200}$, the total path length is
$l\sim 227\ h^{-1}$ kpc at the impact distance of $\rho=76.5\ h^{-1}$
kpc. This path length combined with the inferred column density,
suggests the total hydrogen density is $n({\rm H}) \sim 1.1 \times
10^{-4}\ h$ cm$^{-3}$.  For a uniform density, the total mass in the
hot gas within the virial radius is $M_{\rm gas}\sim 4 \times 10^{10}\
h^{-2}\,M_{\odot}$. This mass is significantly greater than the mass
in stars $M_*\sim 4 \times 10^9\ h^{-2}\,M_{\odot}$ as inferred by the
broad-band spectral energy distribution of the galaxy.  The difference
would suggest that the hot gas revealed by the NeVIII absorption is
the dominant baryonic component.

We have argued based on the mean space density of dwarf galaxies (see
\S\ 4.1) that galaxies $A$ and $B$ share a common halo.  If instead
$A$ is at a cosmological distance from $B$ and responsible for 
the absorption, the
virial temperature is $T_{\rm vir}\sim 2\times 10^5$ K for its
host halo, somewhat lower than the temperature derived by
\citet{savage05}. The gas mass in this case would be 
$M_{\rm gas}\sim 9 \times 10^{9}\
h^{-2}\,M_{\odot}$. This mass is approximately a factor of 10
times larger than the stellar mass, so as is the case for galaxy $B$,
the hot gas would be the dominant baryonic component if the NeVIII
absorption was due to a hot gas halo.

However, the notion of NeVIII arising in an extended hot halo is
difficult to reconcile with the relatively narrow line width observed
for the NeVIII absorption feature, $b_{\rm NeVIII}=23\pm 15$ \kms,
which corresponds to a velocity dispersion of $\sigma_v=16$ \kms.
This is smaller than the velocity dispersion expected from virial
motion along the line of sight $\langle\sigma_v\rangle\equiv v_{\rm
vir}/\sqrt{3}\approx 62 $ km/s and 42 km/s in a uniformly distributed
hot gas in a halo of $M_h=10^{11.5}\ h^{-1}\,M_{\odot}$(B) and
$M_h=10^{11}\ h^{-1}\,M_{\odot}$(A), respectively. We therefore do not
favor this scenario.

\subsection{Photo-ionized Halo Gas}

Next, we consider the possibility that the NeVIII absorber
originates in individual clouds that are photo-ionized by both the
intergalactic radiation field and local ionizing sources.  Recall that
\citet{savage05} ruled out a photo-ionization origin of the NeVIII
absorber because the implied path length given a mean intergalactic
radiation field of $J_{912}=1.9\times 10^{-23}$ ergs cm$^{-2}$
s$^{-1}$ Hz$^{-1}$ sr$^{-1}$ at $z=0.2$ was $\sim$ 11 Mpc, too large
for the observed line widths.  The required path length would be
smaller if the absorber was exposed to a more intense radiation source
such as a local AGN.  Here we examine whether our galaxy data can rule
out the presence of such an ionizing source for producing the NeVIII
absorber.

To reduce the path length to a more reasonable size ($\sim$ 100 kpc)
requires the intensity of the radiation to increase by approximately
two orders of magnitude (up to $\sim$ 2 $\times$ 10$^{-21}$ erg
cm$^{-2}$ s$^{-1}$ Hz$^{-1}$ sr$^{-1}$).  Assuming a flat $f_\nu$
spectrum (appropriate for an AGN) and adopting a $r^{-2}$ scaling law,
we estimate that an AGN would need to be as luminous as
$M_R-5\log\,h=-17.5$ at 38 kpc ($A$) or $M_R-5\log\,h=-19.8$ at
110 kpc
($B$) away in order to provide sufficient ionizing
photons for producing the NeVIII absorber\footnote{Here we adopt $H_0=70$ km s$^{-1}$ Mpc$^{-1}$ for
evaluating the distances between the absorber and galaxies $A$ and
$B$.}.  The expected AGN
luminosity already exceeds the observed luminosity for either of the
two galaxies.  In addition, there is a lack of AGN spectral features
in the optical spectra.  We therefore conclude that galaxies $A$ and
$B$ are unlikely to host a powerful enough AGN to produce the observed
absorption with photo-ionization.

While photo-ionization cannot explain the observed NeVIII absorber, we
note that additional ionizing photons from galaxies $A$ and $B$ may
contribute substantially to the ionizations of H$^0$, C$^{2+}$,
O$^{2+}$, and Si$^{2+}$.  Adopting a flat UV spectrum for star-forming
galaxies and assuming 2\% escape fraction of ionizing photons, we
estimate the absorbing cloud is illuminated by additional ionizing
radiation intensities of $J_{912}=1.3\times 10^{-22}$ and $0.3\times
10^{-22}$ ergs cm$^{-2}$ s$^{-1}$ Hz$^{-1}$ sr$^{-1}$ from galaxies
$A$ and $B$, respectively.  The inferred gas density and size of the
H\,I absorbing cloud are then $n_{\rm H}=2.2\times 10^{-4}$ cm$^{-3}$
and $\approx 4.7$ kpc. By assuming a spherical cloud, we obtain a mass
of ${\rm M}_{cl}\approx 6\times 10^5\,{\rm M}_\odot$. However, the
mass could be larger with other assumed geometries.

\subsection{Outflows}

Next, we consider the possibility that the absorption arises in a
starburst driven wind from the dwarf galaxy ($A$).  Based on
the optical spectrum presented in Figure 1, we find that galaxy $A$ is
forming stars at a rate ($\sim$ 0.1 M$_{\odot}$ yr$^{-1}$) that is
typical of dwarf galaxies at low redshift.  With a rest-frame
H$\alpha$ equivalent width of ${\rm EW}\approx 13$ \AA, however, the
galaxy would not be considered a starburst by most definitions adopted
in the literature \citep{lee08}.  Combining the estimated total
stellar mass for galaxy $A$ of $M_*\sim 8 \times 10^8\
h^{-2}\,M_{\odot}$ and the observed star formation rate, we derive a
characteristic star formation time scale of $\tau\approx 8$ Gyr.  This
is much longer than the characteristic time scale for typical
starburst galaxies \citep{b04}.  We also note that the major axis of
the galaxy is oriented along the direction to the QSO. 
As an outflowing wind is expected to be predominantly along the
minor axis, the bulk of the wind is unlikely to pass in front of the
QSO sightline for the observed geometry.  Given the modest star
formation rate in galaxy $A$ and the disk alignment, a starburst
driven wind would seem incapable of producing the observed absorption
26.5 $h^{-1}$ kpc from the disk of the galaxy.

\subsection{Conduction Front}

Finally, we consider the idea that the NeVIII absorption is produced
at the turbulent boundary layers of cooler clouds moving at high speed
through a hot medium.  This scenario has been proposed to explain the
OVI absorption associated with high velocity clouds and the Magellanic
Stream in the Milky Way \citep{sembach03}.  The redshift difference
between galaxy $B$ and the NeVIII absorber leads to a relative
velocity of nearly 200 \kms, roughly twice the sound speed for an
ionized warm-hot medium of $T\sim 5\times 10^5$ K.  Although the
clouds are expected to suffer from the Kelvin-Helmholtz instability,
the inferred gas density and cloud size discussed in \S\ 4.3 suggest
that the cloud may be massive enough to counter-balance such
instabilities \citep{mb04}

A natural consequence of this scenario is that both low and high
ionization species would be expected with a moderate velocity offset
between the two components ($\sim$ 20 km s$^{-1}$), similar to what is
observed in the absorber at $z=0.20701$.  Therefore, the conductive
front idea is both consistent with the observed galactic environment
and the multi-phase nature of the gas.

\section{Conclusions}

Combining the results of our galaxy survey with the absorption line
analysis of \citet{savage05}, we have considered many possible origins
for the $z=0.20701$ absorber toward HE0226$-$4110.  We find that the
absorber is best explained by a cool cloud moving through a hot
medium.  Although our study suggests that the NeVIII absorption may
not be directly probing the WHIM in this system, it does imply the
existence of hot gas at least 76.5 $h^{-1}$ kpc from a $0.25\,L_*$
galaxy ($B$).  The existence of extended hot gas halos around disk
galaxies is a key ingredient in most semi-analytic models of disk
galaxy formation \citep{wf91,cole00,bower06}. However, there has been
very little direct observational evidence for such halos in quiescent
galaxies. Perhaps the best evidence for an extended hot corona comes
from the Milky Way, where the presence of hot gas has been inferred
from many different observations \citep{sembach03,fang07,bl07}.
Our
observations provide evidence for a similar halo around
galaxy $B$. Given the large extent of the halo, a significant fraction
of the baryons in the system are likely contained in hot gas.

\acknowledgments

We acknowledge valuable comments from the referee and useful
discussions with Nick Gnedin, Juna Kollmeier, Janice Lee, Nicolas
Lehner, Jesper Rasmussen, Josh Simon, Chris Thom and Rik Williams.
H.-W.C. acknowledges partial support from NASA grant
NNG06GC36G and NSF grant AST-0607510.


\clearpage




\end{document}